\documentclass[Journal]{IEEEtran}

\usepackage{setspace}
\usepackage{amsmath,url}
\usepackage{epsfig,amssymb,amsbsy,verbatim,array,enumerate}
\usepackage{pstricks,psfrag,theorem,cite,paralist,subfigure,empheq}
\usepackage{hyperref}
\usepackage{bm}
\usepackage{graphicx}
\usepackage{cite}
\usepackage{url}
\usepackage{color}

\let\intern=\iftrue

\def\figref#1{Fig.\,\ref{#1}}%
\def\E{\mathbb{E}}
\def\P{\mathbb{P}}
\def\R{\mathbb{R}}

\def\ie{{\em i.e.}}

\def\erfc{\operatorname{erfc}}

\def\var{\operatorname{var}}

\def\sir{\mathsf{SIR}}

\def\dd{\mathrm{d}}

\def\SIR{\mathsf{SIR}}
\def\sir{\mathsf{SIR}}

\def\ps{p_{\rm s}}
\def\Ps{P_{\rm s}}

\def\B{{\rm B}}

\def\eqa{\stackrel{{\rm (a)}}{=}}
\def\eqb{\stackrel{{\rm (b)}}{=}}
\def\eqc{\stackrel{{\rm (c)}}{=}}
\def\appa{\stackrel{{\rm (a)}}\approx}
\def\Phib{\Phi_{\rm b}}
\def\Phiu{\Phi_{\rm u}}

\def\PhiI{\Phi_{\rm u}\backslash\{x_0 \}}

\newtheorem{theorem}{Theorem}

\newtheorem{corollary}{Corollary}

\newlength{\figwidth}


\usepackage{xcolor,tikz}

\let\arxiv\iftrue

\begin{document}
\title{The Meta Distribution of the SIR for \\Cellular Networks with Power Control}

\author{Yuanjie~Wang,
        Martin~Haenggi,~\textit{Fellow,~IEEE},
        and Zhenhui~Tan,~\textit{Member,~IEEE}
\thanks{Yuanjie~Wang and Zhenhui~Tan are with the State Key Laboratory of Railway Traffic Control and Safety, Beijing Jiaotong University, Beijing, 100044, China (e-mail: wang.yuanjie@outlook.com). Martin~Haenggi is with the Dept. of Electrical Engineering, University of Notre Dame, IN, 46556, USA (e-mail: mhaenggi@nd.edu). The work was supported by National Natural Science Foundation of China (61471030), National Science and Technology Major Project of China (2015ZX03001027-003) and the US National Science Foundation (grant CCF 1525904).}
}

\maketitle
\IEEEpeerreviewmaketitle

\begin{abstract}
	The \textit{meta distribution} of the signal-to-interference ratio (SIR) provides fine-grained information about the performance of individual links in a wireless network. This paper focuses on the analysis of the meta distribution of the SIR for both the cellular network uplink and downlink with fractional power control. For the uplink scenario, an approximation of the interfering user point process with a non-homogeneous Poisson point process is used. The moments of the meta distribution for both scenarios are calculated. Some bounds, the analytical expression, the mean local delay, and the beta approximation of the meta distribution are provided. The results give interesting insights into the effect of the power control in both the uplink and downlink. Detailed simulations show that the approximations made in the analysis are well justified.	
\end{abstract}


\section{Introduction}
\subsection{Motivation and the meta distribution}
\IEEEPARstart THE META distribution (of the SIR) has been proposed in \cite{MHmeta}, where it is applied to both Poisson bipolar networks and downlink cellular networks without power control to answer the questions such as ``What fraction of users in a network can achieve a desired link reliability given the required SIR threshold?'' or ``How is the success probability of individual links distributed in each network realization?'' Such questions are often asked by network operators and indeed of great significance for providing guidance to the practical deployment of the wireless networks. However, these questions have not been answered analytically for the uplink in cellular networks. The uplink, compared to its downlink counterpart, is more complex in the network structure and requires power control to maintain the link quality and mitigate the inter-cell interference. These differences make the analysis of uplink in the framework of the meta distribution more challenging. Meanwhile, since the meta distribution provides more fine-grained information including the variance of the conditional success probability and the mean local delay, etc., an examination of the effect of power control in the downlink from the perspective of the meta distribution is also warranted. 

Formally, the meta distribution of the SIR in cellular network is defined as
\begin{equation}
\bar F(\theta,x)\triangleq \bar F_{\Ps}(\theta, x) = \P^{o}(\Ps(\theta)>x),
\label{FPs}
\end{equation}
where $\theta\in\R^+$, $x\in [0,1]$, and $\Ps(\theta)\triangleq \P(\sir > \theta \mid \Phi)$ is the conditional success probability averaged over the  fading and the random activities of the interferers given the point process, $\P^{o}$ denotes the Palm measure of the point process, given an active receiver at the origin, and the SIR is measured at that receiver.

$\Ps(\theta)$ can be interpreted as the reliability, \textit{i.e.}, the success probability of the link in consideration given the SIR threshold $\theta$. The meta distribution corresponds to the fraction of links in each network realization that achieve an SIR of $\theta$ with reliability at least $x$. Often operators are interested in the ``5\% user performance'', which is the performance level that
95\% of the users achieve or exceed. Such information can be directly read out from
the meta distribution, while the traditional standard (mean) success probability analysis provides virtually no information about it.

The $b$-th moment of $\Ps(\theta)$ (with respect to the Palm measure) is defined as 
\begin{equation}
M_b(\theta) \triangleq \E^{o}\left(\Ps(\theta)^b\right),\quad b\in \mathbb{C}.
\label{bthmomentDef}
\end{equation}

By this definition and noting that the random variable $\Ps(\theta)\in \left[0, 1\right]$, we have
\[M_b\left(\theta\right) = \int_0^1x^{b}\dd F_{\Ps}(x) \eqa \int_0^1 bx^{b-1} \bar F_{\Ps(x)}\dd x, \] with (a) following integration by parts. For the standard (mean) success probability $\ps(\theta)\triangleq\P(\sir>\theta)$, we easily obtain $\ps(\theta)\equiv M_1(\theta)$, \ie, 
the first moment of the conditional success probability is the standard success probability, as expected. Also, the variance of the conditional success probability is given by ${\rm Var}(\Ps)=M_2-M_1^2$. The variance quantifies the differences in the user experiences. 

\subsection{Related work}
\label{sec:RW}
In the analysis of wireless networks with randomly deployed nodes based on stochastic geometry, the Poisson point process (PPP) is the most widely used model due to its analytical tractability. The tractability of PPP is a consequence of the independence between the points, which is formally captured by Slivnyak's theorem \cite{haenggi2012stochastic}. It also results in a simple expression for the probability generating functional (PGFL) \cite{haenggi2012stochastic}. Most of the existing studies focus on some performance metric as a function of the SIR at the typical receiver in the network obtained by averaging over the channel fading and the point process by utilizing the Laplace transform of the interference, which is obtained from the PGFL. However, such a spatial average only yields limited information about the individual links. In particular, the (mean) success probability $\ps(\theta)$ is one of the most important performance metric of interest, usually interpreted as the probability that the typical receiver achieves a target SIR threshold. Informally, in stochastic geometry, this ``typical point'' (the typical receiver) origins from a selection procedure in which every point of the point process must have the same chance of being selected, hence, $\ps(\theta)$ is in fact a result of \textit{spatial averaging} over the point process. It only quantifies the overall SIR performance of the networks, but provides no detailed statistical information for the performance of each individual link in the network. To obtain fine-grained information on the SIR and reveal how the success probabilities are distributed among the individual links, the meta distribution of the SIR has been formally proposed in \cite{MHmeta}. 

In \cite{MHmeta} the meta distribution is applied to study Poisson bipolar networks with ALOHA channel access and the downlink of Poisson cellular networks, and closed-form expressions of the $b$-th moment of the conditional success probability for the two network classes are derived. The concept of the conditional success probability dates back to \cite{baccelli2010new}, where it is used to study the local delay in ad hoc networks modeled by PPPs. 
In \cite{GantiAndrews2010} the conditional success probability is used to study Poisson bipolar networks without a channel access scheme. The asymptotic behavior of the distribution is studied and the moments are calculated. 
In \cite{2017mHaenggiTcom} the meta distribution is analyzed for the scenario of D2D communication underlaid with the downlink of Poisson cellular networks with ALOHA channel access, and the moments of the conditional SIR distribution, the mean local delay of both the typical D2D receiver and the typical cellular receiver are derived. Another application of the meta distribution is found in \cite{SMHaenggiIcc17}, where the spatial outage capacity, defined as the maximum density of concurrently active links with link reliability greater than a threshold, is proposed and studied for Poisson bipolar networks. Recently, mm-wave D2D networks have also been examined under the framework of the meta distribution in \cite{Deng17}. 

For Poisson downlink cellular networks, the standard assumption is to ignore the fact that there is at most one active user per resource block per cell and condition on a user at the origin, independently of the point process\footnote{As a result, the performance of the typical user corresponds to the average performance of a user population where the mean number of users in each cell is proportional to the cell area.}. This, together with nearest-BS association makes the downlink easy to model and analyze if there is no power control, since the interference at the typical user does not depend on the link distances in other cells, and the interfering BSs form a PPP conditioned on the distance of the typical link. Furthermore, the average interfering signal power from each interfering BS is by definition smaller than the average signal power from the serving BS, which simplifies the derivations. These advantages make the downlink cellular network model thoroughly studied \cite{andrews2011tractable, blaszczyszyn2013using, elsawy2013stochastic, lu2015stochastic, dhillon2012modeling} and easier to be combined with many types of emerging techniques, \textit{e.g.}, cooperative transmission \cite{nigam2014coordinated, nigam2015spatiotemporal, 2016arXiv161001897R}, MIMO \cite{dhillon2013downlink, bjornson2016deploying, george2016ergodic}, and D2D communication \cite{sun2015d2d, liu2015stochastic, ye2014tractable}. 

The uplink case is quite different: on the one hand, the interfering user in a neighboring cell can be much closer to a BS than the transmitting user in the cell of that BS; on the other hand, the interfering user point process seen at the typical BS is not a PPP and hard to model due to the correlation between the Voronoi cells of the \textit{Poisson-Voronoi tessellation} and the channel access scheme in each cell. Moreover, due to the different distance ratios between the desired and interfering links compared to the downlink, it can be expected that power control at the user leads to improved performance in the uplink. 

For the uplink, various models have been proposed to approximate the network performance. \cite{elsawy2014stochastic} analyzes the coverage performance in $K$-tier uplink Poisson cellular networks with truncated channel inversion power control, but the approximation of the interfering user point process by a homogeneous PPP is not accurate since it does not capture the correlations between the interfering points and the BSs under consideration. \cite{lee2014uplink} studies the uplink SIR distributions in a two-tier heterogeneous cellular network by approximating the interferer locations of a tier as a non-uniform Poisson point process whose intensity at a location $x$ is the intensity of the BSs in that tier multiplied by a probability factor that if there was a point of the active user process at $x$, it would belong to the Voronoi cell of another BS (of the same tier) rather than the reference one. A similar non-uniform PPP approximation is also used in \cite{Singh2015TWC} and \cite{2016arXiv160403183A} by conditioning on a user at the origin to analyze the full-load case (\ie, all BSs are active). \cite{7511710} models the interfering user point process as a homogeneous PPP excluding the ball centered at the target BS with the radius determined through matching the average number of excluded points from that homogeneous PPP and the non-homogeneous PPP in \cite{Singh2015TWC}. 

The approximations made in \cite{lee2014uplink, Singh2015TWC, 2016arXiv160403183A, 7511710} are not very accurate, and some of the modeling assumptions are inconsistent, as we will discuss in Remark 1 in Section II.B. These shortcomings are addressed in this paper. 

\subsection{Contributions}
The paper makes the following contributions:
\begin{itemize}
	\item We derive approximate analytical expressions of the $b$-th moment for the Poisson cellular networks with fractional power control for both uplink and downlink and calculate the analytical meta distribution of the SIR for the two scenarios.
	\item We investigate the effect of the fractional power control on the mean local delay, which is the -1-st moment of the conditional success probability.
	\item We show that the meta distribution of the SIR for both the uplink and downlink Poisson cellular networks can be accurately approximated by the beta distribution through matching the first and second moments. 
	\item We reveal the trade-off between the first moment of the conditional success probability and its variance in the medium-$\theta$ regime (from $-10$ dB to $10$ dB) and discuss the optimal operating range of the power control exponent. 
\end{itemize}

\section{Meta Distribution for the Uplink}
\subsection{Network model}
We consider a single-tier uplink Poisson cellular network of type I in \cite{2016arXivUserHaenggi}, where base stations (BSs) are modeled as $\Phib=\Phi\cup\{o\}$, where $\Phi\subset\R^2$ is a homogeneous PPP with intensity $\lambda$. By Slivnyak's theorem, the BS at the origin becomes the typical BS under expectation over $\Phib$. Orthogonal access like OFDMA is assumed, \ie, each BS schedules only one user on each resource block (RB). For a given RB, each user is uniformly distributed in the Voronoi cell of the serving BS. Formally, the user point process is defined as $\Phi_{\rm u} \triangleq \left\lbrace y\in\Phib: U(V(y))\right\rbrace $, where $V(y)$ denotes the Voronoi cell of BS $y$; $U(B),~B\subset\R^2$, denotes a point chosen uniformly and randomly from $B$, and independently across different $B$. 

\begin{figure} [t]
	\begin{center}
		\includegraphics[width=\figwidth]{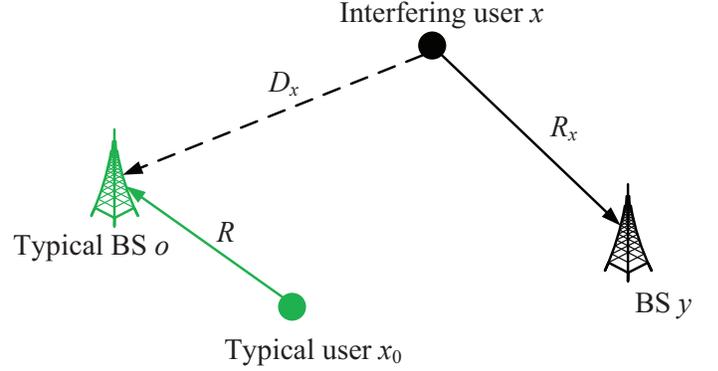}
		\caption{Uplink network model. The typical BS is at the origin, the typical user served by it is $x_0$.}
		\label{fig:ULSysModel}
	\end{center}
\end{figure}

The network model is depicted in \figref{fig:ULSysModel}. The user served by the typical BS is the typical user, and its location is denoted by $x_0$. The distance from each user $x$ to its own serving BS is denoted by $R_x$; for the typical user, the subscript $x_0$ is omitted, so $R = \|x_0\|$. We denote the interfering user point process as $\Phi_{\rm I}$, given by $\Phi_{\rm I} = \PhiI$. The distance from the interfering user located at $x$ to the typical BS is denoted by $D_x = \|x\|$. The standard power-law path loss model with exponent $\alpha > 2$ for signal propagation and the standard Rayleigh fading are used. In this case, the power fading coefficients $h_{x}$ associated with the user at $x$ and the typical BS are exponentially distributed variables with unit mean, \ie, $h_{x}\sim\exp(1)$. We assume $\{h_{x}\}$ are independent for all $x\in\Phi_{\rm u}$.     

We use fractional power control at the user in the form $P_x = p_0R_x^{\alpha\epsilon}$, which is one of the most widely used schemes for the uplink cellular networks. The power control exponent $\epsilon\in[0,1]$ is introduced to partially compensate for the path loss, $p_0$ is the baseline transmit power when there is no power control. Noise is neglected, \ie, an interference-limited scenario is considered.

For our network setup, the uplink SIR at the typical BS (on a given RB) is given by 
\[\SIR \triangleq \frac{P_{x_0} h_{x_0}R^{-\alpha}}{\sum\limits_{x \in \Phi_{\rm I}}P_x h_{x} D_x^{-\alpha}}.\]

Thus, for the typical user, $P_{x_0} = p_0R^{\alpha\epsilon}$,
\begin{equation}\label{eq:UplinkSIR}
\SIR = \frac{h_{x_0}R^{\alpha(\epsilon-1)}}{\sum\limits_{x \in \Phi_{\rm I}}R_x^{\alpha \epsilon} h_{x} D_x^{-\alpha}}.
\end{equation}

To analyze the SIR performance, we need to know the statistical properties of $\Phi_{\rm I}$ and the distribution of the link distances in the network.

\subsection{Approximations of interferer process and link distance distribution}
\label{sec:PPPApprx}
An exact analysis of the network seems unfeasible, hence there is a need for sensible approximations. Since $o\in \Phib$ and the typical user is not an interferer, $\Phi_{\rm I}$ is a non-stationary process with intensity depending on the distance from the origin and whose exact statistics are hopeless to derive. We use the approximations provided in \cite{2016arXivUserHaenggi} for the intensity function of $\Phi_{\rm I}$ and the probability density function (pdf) of the distribution of the link distance $R$, given by
\begin{equation}\label{eq:new}  
\lambda_{\rm I}(x)=\lambda\left(1-\exp\left((-12/5)\lambda\pi \|x\|^2\right)\right), 
\end{equation}
\begin{equation}\label{eq:Rdistr2} 
f_R(r) = \frac{5}{2}\pi\lambda r\exp\left(-\frac{5}{4}\lambda\pi r^2\right), ~~r\geq0.
\end{equation} 

The link distances $R_x$ in the interfering cells are identically distributed as $R$, since all cells are statistically the same. However, they are not independent since the areas and shapes of neighboring cells are correlated, and $R_x$ cannot be larger than $D_x$, hence we characterize the distribution of $R_x$ by conditioning on $D_x$. This results in the truncated Rayleigh distribution
\begin{equation}\label{eq:pdfRx2}
f_{R_x}(r\mid D_x) = \frac{(5/2)\pi\lambda r\exp(-(5/4)\lambda\pi r^2)}{1-\exp(-(5/4)\lambda\pi D_x^2)},  ~~0\leq r\leq D_x.
\end{equation}

This truncation reflects the correlation between $R_x$ and $D_x$ that is present for $x$ near the origin. Equipped with \eqref{eq:new}, \eqref{eq:Rdistr2} and \eqref{eq:pdfRx2}, we can derive an approximation of the $b$-th moment of $\Ps(\theta)$.

\textbf{Remark 1}: The model used here is similar but the approximations have important differences to those used by the previous works \cite{elsawy2014stochastic}, \cite{Singh2015TWC, 2016arXiv160403183A, 7511710}, which also consider the full-load uplink scenario and single-user scheduling on each resource block. Their network model is in accordance with the type II model\footnote{The type II user point process is defined as $ \Phiu\triangleq\{y\in\Phib\colon U(V(y)\cap\Phi_0)\}$ and $\Phi_0$ denotes the entire user population which is a point process independent from $\Phib$. The users in each cell are chosen only from the countable set $V(y)\cap\Phi_0,~y\in\Phib$, hence the number of users in a cell can be more than one, but can also be zero.} in \cite{2016arXivUserHaenggi}, but they argue that their type II model is identical to a type I model. What they are actually analyzing, though, is neither---they are analyzing a model where the users are placed independently of the BSs, resulting in the performance of a user in the \textit{Crofton cell} \cite{PCalkaCrofton}. Hence, their ``typical user'' is not, in fact, the typical active user.
 
\begin{figure} [t]
	\begin{center}
		\includegraphics[width=\figwidth]{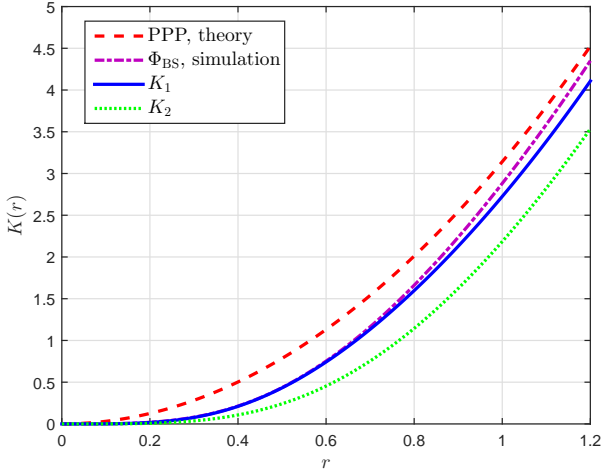}
		\caption{Ripley's \textit{K} function. $K_1(r) = \pi r^2 + 5/12 e^{-12/5 \pi r^2}-5/12 $ corresponds to (5) in \cite{2016arXivUserHaenggi} and $K_2(r) = \pi r^2 + e^{-\pi r^2}-1$ is \textit{K} function of the interfering user point process approximated by the non-homogeneous PPP in \cite{Singh2015TWC} and \cite{2016arXiv160403183A}. Both \textit{K} functions of the standard PPP in theory and that of the interfering user point process at the typical BS by simulation are shown for comparison. }
		\label{fig:Kfun}
	\end{center}
\end{figure}

Further, \cite{Singh2015TWC} and \cite{2016arXiv160403183A} approximate $\Phi_{\rm I}$ by a non-homogeneous PPP with intensity 
\begin{equation}\label{eq:old} 
\lambda_{\rm I}(x)=\lambda(1-\exp(-\lambda\pi \|x\|^2)), 
\end{equation}
with the factor $1-\exp(-\lambda\pi \|x\|^2)$ being interpreted as the probability that a user at the point $x$ is interfering to the typical BS at the origin. However, wherever a user not served by the reference BS is located, it is for sure an interferer, so this probability is 1, trivially. Such an approximation does not accurately reflect the pair correlation between the BS and the points in $\Phi_{\rm I}$. We can verify this by inspecting Ripley's \textit{K} function in \figref{fig:Kfun}. 
It clearly shows that the approximation \eqref{eq:old} for the uplink interfering user point process underestimates the number of the interferers in the proximity of the BS (\ie, in the range of $r$ from 0.4 to 1), thereby underestimating the average aggregate interference suffered by the typical BS, while the approximation \eqref{eq:new} closely matches the simulation result. We can also see that the interfering user point process outside a large enough radius of the typical BS approaches the PPP asymptotically and the new approximation \eqref{eq:new} also slightly underestimates the $K$ function, but this has negligible effect on the interference as the signal power decays quickly with the distance.     

Next, as a result of the Crofton cell, the distribution for the link distance $R$ in \cite{Singh2015TWC} and \cite{2016arXiv160403183A} is the standard Rayleigh distribution
\begin{equation}\label{eq:Rdistr} 
f_R(r) = 2\pi\lambda r\exp(-\lambda\pi r^2), ~~r\geq0.
\end{equation}

\begin{figure} [t]
	\begin{center}
		\includegraphics[width=\figwidth]{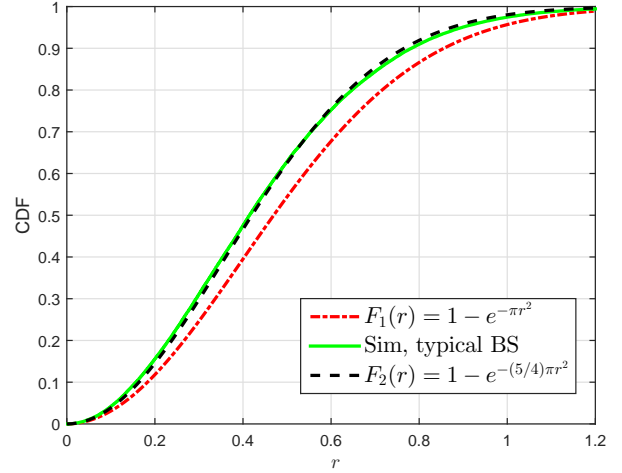}
		\caption{Link distance distribution comparison. $F_1(r)$ corresponds to the pdf in \eqref{eq:Rdistr}, $F_2(r)$ corresponds to the pdf in \eqref{eq:Rdistr2}, both for $\lambda = 1$.}
		\label{fig:DistComp}
	\end{center}
\end{figure}

However, it is easily seen from \figref{fig:DistComp} that \eqref{eq:Rdistr} deviates significantly from the actual link distance distribution obtained from the simulation. 

The approximations and analyses in this paper overcome these shortcomings of prior work: the inconsistency in the modeling of the users and, as a result, between Crofton and typical cells, the link distance distributions, and the inaccuracy in the intensity function of the interfering point process.

\subsection{Moments}
\begin{theorem}[Moments for the uplink with FPC]
	The $b$-th moment $M_b$ of the conditional success probability of the uplink Poisson cellular networks with FPC under the non-homogeneous PPP approximation \eqref{eq:new} for the interfering user point process is closely approximated by
	\begin{equation}
	\tilde{M_b} = \int_{0}^{\infty} \exp\left(-z\left(1+\int_{0}^{\infty}f_b\left(z,x\right) \dd x\right)\right) \dd z,
	\label{eq:bthmoment}
	\end{equation}
	where 
	\begin{align*}
	f_{b}(z,x)=&\displaystyle\int_{0}^{x} B_1^{-1}ze^{-zy}\frac{1-e^{-zxB_2/B_1}}{1-e^{-zx}}\nonumber\\
	           &\times\left(1- \left(1+\theta y^{\frac{\alpha \epsilon}{2}}x^{-\frac{\alpha}{2}}\right)^{-b} \right) \dd y,
	\end{align*}
	$B_1=5/4$, $B_2=12/5$, and $\epsilon \in [0, 1]$.
	\label{bthmoment}
\end{theorem}

\begin{IEEEproof}
	The conditional coverage probability is
	\begin{align}
	\Ps(\theta)&= \P\Big(h_{x_0} > \theta \sum_{x\in\Phi_{\rm I}} h_xD_x^{-\alpha}R_x^{\alpha \epsilon}R^{\alpha(1-\epsilon)} \: \Big| \: \Phi_{\rm a},\Phi_{\rm b}\Big) \nonumber\\
	&=\prod_{x\in\Phi_{\rm I}} \frac{1}{1+\theta \big(\frac{R^{(1-\epsilon)}R_x^{\epsilon}}{D_x}\big)^\alpha} .
	\label{eq:Mb-FPC}
	\end{align}
	
	\allowdisplaybreaks
	
	Then $M_b$ follows as
	\begin{align}
	M_b&=\mathbb{E}\prod\limits_{x \in \Phi_{\rm I}} \frac{1}{\big(1+\theta \frac{R^{\alpha(1-\epsilon)}R_x^{\alpha \epsilon}}{D_x^{\alpha}}\big)^b} \nonumber \\
	&=\mathbb{E}\prod\limits_{x \in \Phi_{\rm I}} \mathbb{E}_{R_x}\bigg(\frac{1}{\big(1+\theta \frac{R^{\alpha(1-\epsilon)}R_x^{\alpha \epsilon}}{D_x^{\alpha}}\big)^b} \mid D_x, R\bigg) \nonumber \\
	&\appa\mathbb{E}\prod\limits_{x \in \Phi_{\rm I}} \int_{0}^{D_x} \frac{(2B_1)\pi\lambda x e^{-B_1\lambda\pi x^2}}{(1-e^{-B_1\lambda\pi D_x^2})\left(1+\theta x^{\alpha\epsilon} D_x^{-\alpha} R^{\alpha\left(1-\epsilon\right)}\right)^b} \dd x \nonumber \\
	&\eqb\mathbb{E}_R\exp\Bigg(\int_0^\infty -2\lambda\pi a\big(1-e^{-B_2\lambda\pi a^2}\big)\Big(1 \nonumber\\
	&~~~~-\int_0^a \frac{2B_1\pi\lambda x e^{-B_1\lambda\pi x^2}}{(1-e^{-B_1\lambda\pi a^2})\big(1+\theta x^{\alpha\epsilon} a^{-\alpha}R^{\alpha\left(1-\epsilon\right)}\big)^b} \dd x\Big) \dd a\Bigg) \nonumber \\
	&\eqc\int_0^\infty 2B_1\pi\lambda r \exp\Bigg(\int_0^\infty -2\lambda\pi\big(1-e^{-B_2\lambda\pi a^2}\big)\bigg(1 \nonumber\\
	&~~~~-\int_0^a \frac{2B_1\pi\lambda x e^{-B_1\pi\lambda x^2}}{(1-e^{-B_1\lambda\pi a^2})\big(1+\theta x^{\alpha\epsilon} a^{-\alpha}r^{\alpha\left(1-\epsilon\right)}\big)^b} \dd x\bigg)\nonumber\\
	&~~~~\times a \dd a \Bigg)e^{-B_1\lambda\pi r^2}\dd r,
	\label{eq:bthmomentexpanded}
	\end{align}
	where (a) uses \eqref{eq:pdfRx2} to average over $R_x$; (b) follows from \eqref{eq:new} and the PGFL of the general PPP \cite{haenggi2012stochastic}; (c) uses \eqref{eq:Rdistr2} to average over $R$.  
	
	Then by using substitution $x/r = u$, $r/x = v$ and $e^{-\lambda \pi r^2} = t$, after some simplification, we obtain $M_b\approx\tilde{M_b}$, with $\tilde{M_b}$ given in \eqref{eq:bthmoment}.
\end{IEEEproof}

\begin{corollary}[Special case: $\epsilon=1$]
	\label{cor:ep1}
	When $\epsilon=1$, 
	\begin{equation} \label{eq:corep1} 
	\tilde{M_b}	= \exp\Big(\int_0^1\int_0^1 \frac{1-u^{B_2/B_1}}{B_1(1-u)}\left(A-u^{x-1}h(x)\right) \dd u \dd x\Big),
	\end{equation}	
	where $B_1 = 5/4$, $B_2=12/5$, $A = 1-(1+\theta)^{-b}$, and
	\[h(x) = \frac{b\theta\alpha x^{\alpha/2-1}}{2(1+\theta x^{\alpha/2})^{b+1}}.\]
\end{corollary}
\begin{IEEEproof}
	By substituting $z = \ln t$ and $\epsilon=1$ into \eqref{eq:bthmoment}, after some simplification, we obtain $\tilde{M_b}$ in the form	
	\begin{equation}
	\tilde{M_b} = \displaystyle\int_{0}^{1} t^{g(t)} \dd t,
	\label{eq:bthmoment2}
	\end{equation}		
where 
\begin{align}
	g(t) &= \int_{0}^{\infty} \int_{0}^{s} -\frac{1}{B_1}\frac{1-t^{sB_2/B_1}}{1-t^s}\ln t \cdot t^y \nonumber\\
	&~~~~\times\left(1-\left(1+\theta y^{\frac{\alpha}{2}}s^{-\frac{\alpha}{2}}\right)^{-b} \right) \dd y \dd s. 
\end{align}
$g(t)$ can be expressed as 
	\begin{align}
	g(t)\eqa& -\ln t \int_{0}^{\infty} \frac{1}{B_1}\frac{1-t^{sB_2/B_1}}{1-t^s} \int_{0}^{1} st^{sx}\nonumber\\
	&~~~~\times\bigg(1-\frac{1}{\big(1+\theta x^{\frac{\alpha}{2}}\big)^{b}} \bigg)\dd x \dd s \nonumber\\
	\eqb& -\int_{0}^{\infty} \frac{1}{B_1}\frac{1-t^{sB_2/B_1}}{1-t^s} \left(\int_0^1\left(At^s-t^{sx}h(x)\right)\dd x\right) \dd s \nonumber\\ 
	\eqc& \frac{1}{B_1}\log_te\int_0^1\int_0^1 \frac{1-u^{B_2/B_1}}{1-u}\left(A-u^{x-1}h(x)\right) \dd u \dd x, 
	\end{align} 
	where (a) follows from the substitution $y/s = x$; (b) from integration by parts; (c) from replacing $t^s$ with $u$. Finally, by inserting $g(t)$ into \eqref{eq:bthmoment2} we obtain \eqref{eq:corep1}.    
\end{IEEEproof}

\begin{figure} [t]
	\begin{center}
		\includegraphics[width=\figwidth]{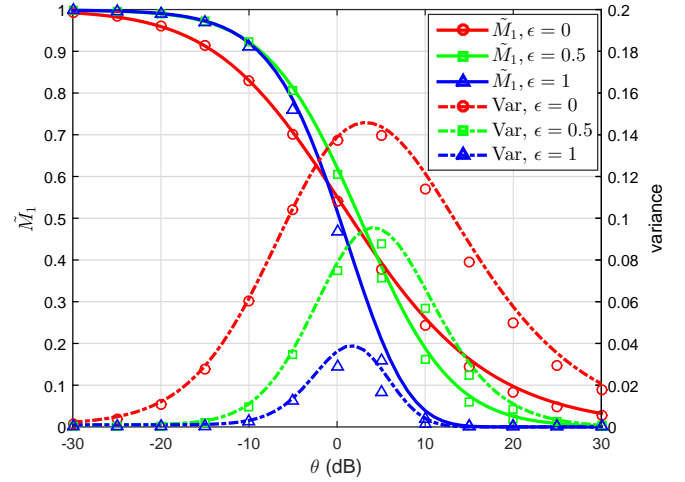}
		\caption{The first moment and variance of $P_s(\theta)$ in the uplink for different $\epsilon$, $\alpha=4$. The curves are the analytical results from Theorem \ref{bthmoment}; the markers correspond to the simulation results.}
		\label{fig:UL-M1-Var}
	\end{center}
\end{figure}

\figref{fig:UL-M1-Var} shows the standard success probability $p_{\rm s} = M_1$ and the variance of the conditional success probability as a function of $\theta$ for $\alpha = 4$ and FPC parameter $\epsilon = 0, 0.5, 1$. The solid and dashed curves correspond to the simulation results, the markers correspond to the analytical results in Theorem \ref{bthmoment}. These results reveal how the FPC parameter affects $M_1$ and the variance. For the low-$\theta$ regime, a higher $\epsilon$ benefits both $M_1$ and the variance, which means it improves the average performance of the network and also reduces the difference between the individual links, resulting in better fairness. Conversely, a higher FPC exponent $\epsilon$ harms the average performance of the network in the high-$\theta$ regime, because for large $\theta$, assuming no power control, only users very close to their BSs will succeed. However, with power control, these users have to drastically reduce their transmit powers relative to the interferers, which reduces the received signal strength at their BSs and thus the SIR. Hence, they are increasingly less likely to succeed as $\epsilon$ grows. The users far away from their BSs are unlikely to benefit from the path loss compensation due to the high SIR threshold. As a result, the average network performance is brought down. 

We can also observe from \figref{fig:UL-M1-Var} that in the medium-$\theta$ regime, there is a trade-off between $M_1$ and the variance. In particular, with $\theta\in [-10, 10]$~(dB), the variance first increases and then decreases while $M_1$ is monotonically decreasing, so there is a compromise between optimizing $M_1$ and the variance by choosing the parameter $\epsilon$. In the high-$\theta$ regime (\ie, the low-reliability regime), $M_1$ and the variance are both monotonically decreasing, while in the low-$\theta$ regime, a better $M_1$ is always accompanied with a lower variance.

\figref{fig:UL-M1-Var} also shows the significance of the meta distribution as a much more refined metric than just $M_1$. For example, if the target SIR is -3 dB, all values of $\epsilon$ lead to a very similar $M_1$, so $M_1$ alone does not tell us which $\epsilon$ to use, but if we consider the variance, it is evident that $\epsilon=1$ is best. 

\begin{figure} [t]
	\begin{center}
		\includegraphics[width=\figwidth]{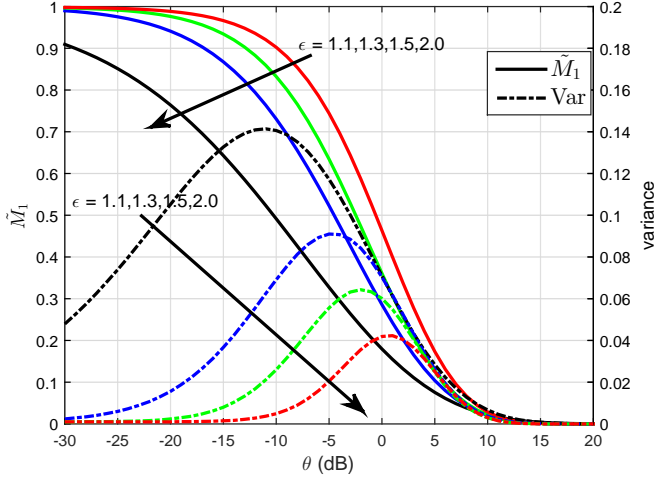}
		\caption{Uplink $\tilde{M_1}$ and variance of $P_s(\theta)$ obtained from \eqref{eq:bthmoment} for different values of $\epsilon$ and $\alpha=4$.}
		\label{fig:M1VarOverCompensation}
	\end{center}
\end{figure}

Theoretically, $\epsilon$ can be greater than 1, which is the over-compensation case. The analytical results are shown in \figref{fig:M1VarOverCompensation}. It is shown that generally speaking, over-compensation of the transmit power has no benefit to the network, especially for the high-SIR threshold regime. This is because raising the transmit power too much makes the receiver more likely to experience interference from transmitters far away, and the incremental interference power dominates the increment of the useful signal power. 

\begin{corollary}[Asymptotic property of $\bm{\epsilon_{\rm{opt}}}$ for the uplink]
	\label{cor:AsymptoticSIRInfty}
	Define $\epsilon_{\rm{opt}}^{(1)}(\theta) = \arg\max\limits_{\epsilon}\tilde{M_1}(\theta)$, $\theta\in\R^{+}$. Then we have 
    $\lim\limits_{\theta\to\infty} \epsilon_{\rm{opt}}^{(1)}(\theta) = 0$.
\end{corollary}
\begin{IEEEproof}
	By observing the structure of the expression \eqref{eq:bthmoment} for $b=1$, since the exponential function is monotonically increasing, and the integration interval is the positive real axis, it is easy to see that to maximize $\tilde{M_1}$,  $-1-\int_{0}^{\infty}f_{1}\left(z,x\right) \dd x$ should reach the maximum. Then the integrand $f_{1}(z,x)=\int_{0}^{x} ze^{-zy} \big(1- (1+\theta y^{\epsilon \alpha/2}x^{-\alpha/2})^{-1} \big) \dd y$ should assume its minimum since it is always positive. For the integrand of $f_{1}(z,x)$, the factor $e^{-zy}$ is monotonically decreasing for $y\in(0,x)$, while the factor $1- \left(1+\theta y^{\epsilon \alpha/2}x^{-\alpha/2}\right)^{-1} = \frac{y^{\epsilon \alpha/2}x^{-\alpha/2}}{\theta^{-1}+y^{\epsilon \alpha/2}x^{-\alpha/2}}$, which is monotonically increasing for $y\in(0,x)$, when $\theta\to\infty$, $\frac{y^{\epsilon \alpha/2}x^{-\alpha/2}}{\theta^{-1}+y^{\epsilon \alpha/2}x^{-\alpha/2}}$ will approach 1, the dominant term is $e^{-zy}$. Thus the integral is minimized at $\epsilon = 0$.    
\end{IEEEproof}

\begin{figure} [t]
	\begin{center}
		\includegraphics[width=\figwidth]{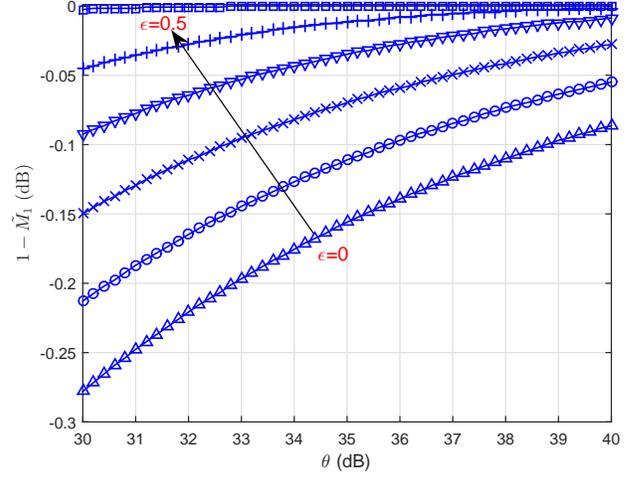}
		\caption{The uplink outage $1-\tilde{M_1}$ for large $\theta$ and $\epsilon = 0, 0.1, 0.2, 0.3, 0.4, 0.5$ from bottom to top.}
		\label{fig:ULEnvelopeEpsilon}
	\end{center}
\end{figure}

\figref{fig:ULEnvelopeEpsilon} illustrates the result in Cor.~\ref{cor:AsymptoticSIRInfty}. It can be seen that as $\theta$ increases, the outage $1-\tilde{M_1}$ (in dB) for $\epsilon=0$ is always below the curves for the other values of $\epsilon$, which means that for large enough $\theta$, $\tilde{M_1}$ is maximized at $\epsilon=0$.    

\begin{figure} [t]
	\begin{center}
		\includegraphics[width=\figwidth]{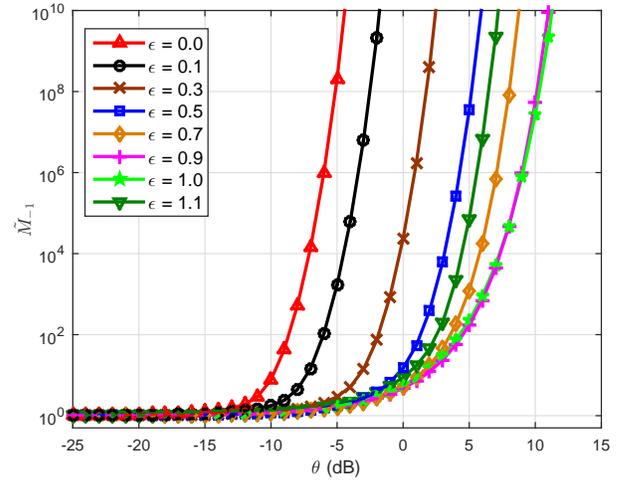}
		\caption{Analytical results of the mean local delay $\tilde{M}_{-1}$ as a function of $\theta$ for different values of $\epsilon$ and $\alpha=3$ in the uplink.}
		\label{fig:MeanLocalDelay_ULal3}
	\end{center}
\end{figure}

\begin{figure} [t]
	\begin{center}
		\includegraphics[width=\figwidth]{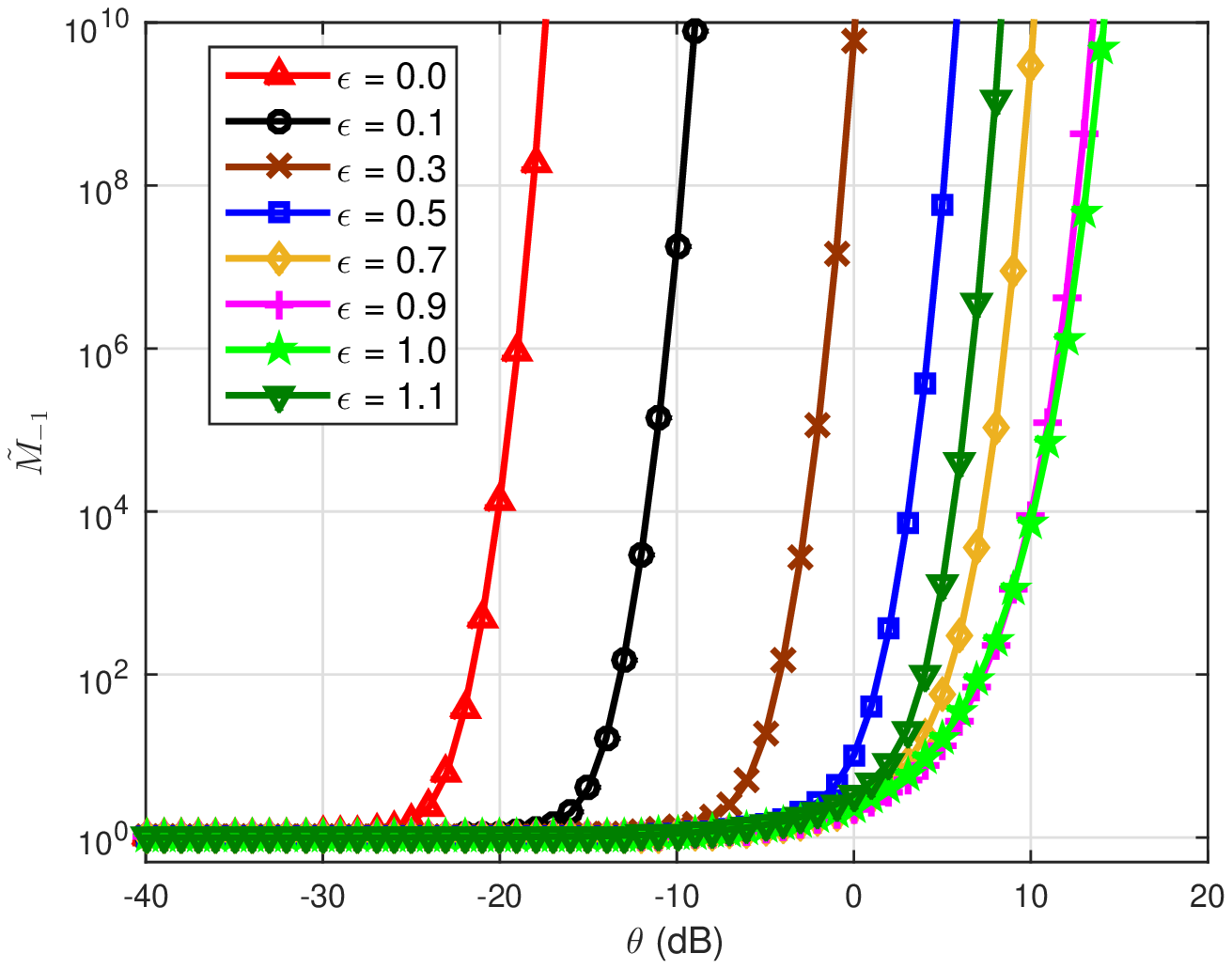}
		\caption{Analytical results of the mean local delay $\tilde{M}_{-1}$ as a function of $\theta$ for different values of $\epsilon$ and $\alpha=4$ in the uplink.}
		\label{fig:MeanLocalDelay_ULal4}
	\end{center}
\end{figure}

$M_{-1}$ is the \textit{mean local delay}, which quantifies the mean number of transmission attempts needed before the first success if the transmitter is allowed to keep transmitting \cite{6353585}. According to \cite{MHmeta}, the mean local delay of downlink Poisson cellular networks without power control exhibits a \textit{phase transition} from finite to infinite when the SIR threshold reaches a critical value. But for uplink Poisson cellular networks with FPC, the curves in \figref{fig:MeanLocalDelay_ULal3} and \figref{fig:MeanLocalDelay_ULal4} show that no phase transition may occur. For a given FPC exponent $\epsilon$, the mean local delay stays close to 1 for small and modest values of $\theta$ and quickly increases at higher $\theta$. A higher $\epsilon$ is helpful in terms of broadening the SIR range for which the mean local delay is below some threshold. An increase past $\epsilon=1$, however, is detrimental.

\textbf{Remark 2:} For $\epsilon=1$, $\tilde{M}_{-1}$ has a simplified expression by Cor.~\ref{cor:ep1}, given by
\begin{align}
&\tilde{M}_{-1} \nonumber\\
&= \exp\Big(-\theta \int_{0}^{1}\int_0^1 \frac{1-u^{B_2/B_1}}{B_1(1-u)}\big(1-\frac{\alpha}{2}u^{x-1}x^{\alpha/2-1}\big)\Big) \dd u \dd x,
\end{align} 
where $B_1 = 5/4,~B_2 = 12/5$. Further, noticing that $B_2/B_1=48/25\approx2$, $\tilde{M}_{-1}$ can be further simplified to the closed-form $M_{-1}=\exp\left(-\theta\cdot C(\delta)\right)$, with $\delta=2/\alpha$ and $C(\delta)$ given by 
\begin{align}
&C(\delta) \nonumber\\
&= \frac{1}{B_1}\Bigg(\frac{3}{2}-\frac{1}{\delta}\bigg(\frac{2\delta}{1-\delta}+\ln 2-\psi\Big(\frac{1-\delta}{\delta}\Big)+\psi\Big(\frac{1-\delta}{2\delta}\Big)\bigg)\Bigg),
\end{align} 
where $\psi(\cdot)=\ln \Gamma(\cdot)$ is the Polygamma function. This shows that no phase transition occurs at $\epsilon=1$. 

\subsection{Meta distribution: analytical expression and classical bounds}
Equipped with a tight approximation for $M_b$, the analytical meta distribution of the SIR for uplink Poisson cellular networks can be obtained from the Gil-Pelaez theorem \cite{gil1951note} as
\begin{equation}
\bar F(\theta,x)\approx\frac12+\frac1\pi\int_0^\infty \frac{\Im(e^{-jt\log x}\tilde{M}_{jt})}{t}\dd t, 
\label{exact}
\end{equation}
where $\Im(z)$ denotes the imaginary part of the complex number $z$.

As in \cite{MHmeta}, some classical bounds can also be directly obtained as follows,

For $x\in [0,1]$, the Markov bounds of the meta distribution are given by 
\begin{equation}
1-\frac{\E^{!t}((1-\Ps(\theta))^b)}{(1-x)^b} < \bar F(\theta,x) \leq \frac{\tilde{M}_b}{x^b} , \quad b>0.
\label{markov}
\end{equation}

For $x\in [0,1]$, let $\tilde V\triangleq \tilde{M}_2-\tilde{M}_1^2$, the Chebyshev bounds of the meta distribution are given by 
\begin{equation}
\bar F_{\Ps}(x)  > 1-\frac{\tilde V}{(x-\tilde{M}_1)^2} , ~~ x<\tilde{M}_1,
\label{cheby1}
\end{equation}
and
\begin{equation}
\bar F_{\Ps}(x)  > 1-\frac{\tilde V}{(x-\tilde{M}_1)^2} , ~~  x>\tilde{M}_1.
\label{cheby2}
\end{equation} 

For $x\in [0,1]$, the Paley-Zygmund bound of the meta distribution is given by 
\begin{equation}
\bar F_{\Ps}(x\tilde{M}_1)\geq \frac{(1-x)^2 \tilde{M}_1^2}{\tilde{M}_2+x(x-2)\tilde{M}_1^2},\quad x\in (0,1).
\label{PZ}
\end{equation}

The Paley-Zygmund bound can be used to roughly quantify the fraction of links that can achieve a certain fraction of the average performance. 

\begin{figure} [t]
	\begin{center}
		\includegraphics[width=\figwidth]{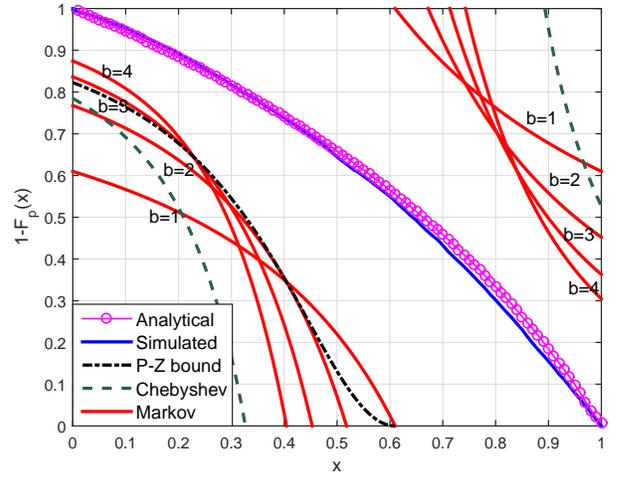}
		\caption{The analytical meta distribution \eqref{exact}, the simulated curve, the Markov bounds \eqref{markov} for $b\in[4]$, the Chebyshev bounds \eqref{cheby1} and \eqref{cheby2} and the Paley-Zygmund bound \eqref{PZ} for $\alpha=4$, $\theta = 0$ dB and $\epsilon=0.5$ in the uplink.  }
		\label{fig:BoundofMeta}
	\end{center}
\end{figure}

\figref{fig:BoundofMeta} shows the meta distribution from both the simulation result and the analytical expression in \eqref{exact}. It verifies the accuracy of the approximation of the moments given in Theorem~\ref{bthmoment}. The classical bounds are also illustrated in this figure. \figref{fig:BoundofMeta} also gives information about the 5\% user performance: in this case, the 5\% user achieves about 10\% reliability.

\begin{figure} [t]
	\begin{center}
		\includegraphics[width=\figwidth]{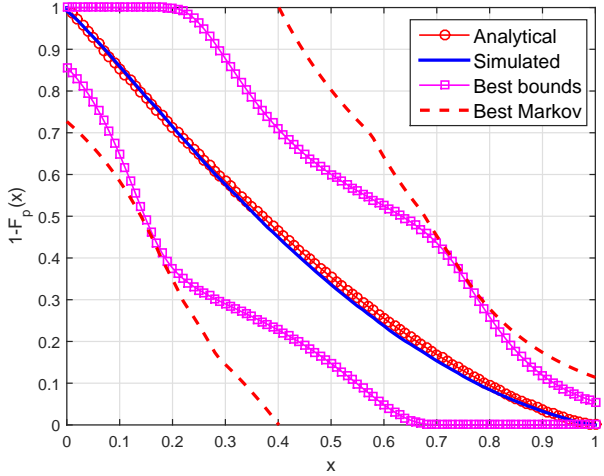}
		\caption{The meta distribution, the best Markov bounds \eqref{markov} for $b\in[4]$, and the best bounds (given the first four moments) for $\alpha=3$, $\theta = 0$ dB and $\epsilon=0.5$ in the uplink.  }
		\label{fig:BestBoundofMeta}
	\end{center}
\end{figure}

Given the first $k$ moments of $P_s(\theta)$, we can establish the tightest possible lower and upper bounds following the procedure in \cite{racz2006moments}, which has also been applied in \cite{MHmeta}.
\figref{fig:BestBoundofMeta} shows the best bounds and the lower and upper envelopes of the Markov bounds for $b \in[4]$.

\subsection{Meta distribution: beta distribution approximation}
Since $\Ps(\theta)$ is supported on $[0,1]$, it is natural to approximate its distribution
with the beta distribution.
The pdf of a beta distributed random variable $X$ with mean $\mu$ is
\[ f_X(x)=\frac{x^{\frac{\mu(\beta+1)-1}{1-\mu}}(1-x)^{\beta-1}}{\B(\mu\beta/(1-\mu),\beta)} ,\]
where $\B(\cdot,\cdot)$ is the beta function.
The variance is given by
\[ \sigma^2\triangleq \var X=\frac{\mu(1-\mu)^2}{\beta+1-\mu} .\]
Matching the mean and variance $\sigma^2$ yields $\mu=\tilde{M}_1$ and
\[ \beta=\frac{(\mu-\tilde{M}_2)(1-\mu)}{\tilde{M}_2-\mu^2}.\]
\begin{figure} [t]
	\begin{center}
		\includegraphics[width=\figwidth]{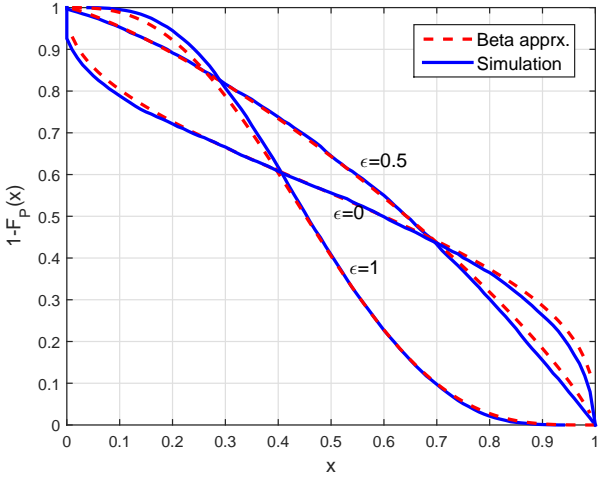}
		\caption{The simulated meta distribution and the beta distribution approximation for $\alpha=4$, $\theta = 0$ dB and $\epsilon \in \{0, 0.5, 1\}$ in the uplink. }
		\label{fig:BetaApprxMeta}
	\end{center}
\end{figure}
~~\figref{fig:BetaApprxMeta} shows that the beta distribution is an almost perfect approximation for the meta distribution. even if the moments are obtained from the approximations $\tilde{M}_1$, $\tilde{M}_2$. The close match between the meta distribution and the beta distribution is very convenient, since it implies that for most purposes, an evaluation of the Gil-Pelaez integral \eqref{exact} is not needed. Also, the beta approximation provides a simple way to help the operator determine the most appropriate parameters for maximizing the performance metrics, e.g., $M_1$ and the variance, or the performance of the 5\% user.

\subsection{Effect of peak power constraint}
In practice, there usually is the maximum transmit power constraint. We can incorporate such a constraint into our analytical framework by considering the truncated fractional power control (TFPC) model
\begin{equation}\label{eq:TFPC}
P_x = 
\begin{cases}
R_x^{\alpha\epsilon}&  R_x \leq \hat{p}^{\frac{1}{\alpha\epsilon}},\\
\hat{p}&  R_x > \hat{p}^{\frac{1}{\alpha\epsilon}},
\end{cases}
\end{equation}
where we assume the transmit power over unit distance is normalized to $1$, without loss of generality. $\hat{p}$ is the (normalized) maximum transmit power, $\alpha$ is path loss exponent and $\epsilon\in[0,1]$ is the path loss compensation exponent, $R_x$ is the link distance in the Voronoi cell of BS $x$. It can easily be seen that when $\hat{p}\to\infty$, TFPC degenerates to FPC. 
\begin{theorem}[Moments for the uplink with TPFC]
	The $b$-th moment of the conditional success probability of the uplink Poisson cellular networks with TFPC is closely approximated by
	\begin{align}
	\tilde{M}_b &= \int_{0}^{S_\epsilon} \exp\Big(-z\big(1+f_{b,1}(z)+f_{b,2}(z)\big)\Big) \dd z \nonumber\\
	&~~~~ +\int_{S_\epsilon}^{\infty} \exp\Big(-z\big(1+g_{b,1}(z)+g_{b,2}(z)\big)\Big) \dd z,
	\label{eq:TFPCULMb}
	\end{align}
	where $S_\epsilon = B_1\lambda\pi \hat p^{\frac{\delta}{\epsilon}}$ and
\begin{figure*}[!t]
	\normalsize
	\begin{align}
	f_{b,2}(z)&=\int_{S_\epsilon z}^{\infty}\int_{S_\epsilon z}^{x} B_1^{-1}ze^{-zy}\frac{1-e^{-zxB_2/B_1}}{1-e^{-zx}}\big[\big(1+\theta y^{\frac{\alpha \epsilon}{2}}x^{-\frac{\alpha}{2}}\big)^{-b} - \big(1+\theta S_\epsilon^{\frac{\alpha \epsilon}{2}} z^{-\frac{\alpha \epsilon}{2}}x^{-\frac{\alpha}{2}}\big)^{-b} \big]\dd y\dd x, \nonumber\\	
	g_{b,2}(z)&=\int_{\frac{S_\epsilon}{z}}^{\infty}\int_{\frac{S_\epsilon}{z}}^{x} B_1^{-1}ze^{-zy}\frac{1-e^{-zxB_2/B_1}}{1-e^{-zx}}\big[\big(1+\theta S_\epsilon^{-\frac{\alpha \epsilon}{2}} z^{\frac{\alpha \epsilon}{2}} y^{\frac{\alpha \epsilon}{2}}x^{-\frac{\alpha}{2}}\big)^{-b} - \big(1+\theta x^{-\frac{\alpha}{2}}\big)^{-b} \big]\dd y\dd x. \nonumber					
	\end{align} 
	\hrulefill
	\vspace*{1pt}
\end{figure*}	

	\begin{align}
	\label{eq:ULTFPC_fg}
	f_{b,1}(z)&=\int_{0}^{\infty}\int_{0}^{x} B_1^{-1}ze^{-zy}\frac{1-e^{-zxB_2/B_1}}{1-e^{-zx}}\nonumber\\
	&~~~~\times\big[1-\big(1+\theta y^{\frac{\alpha \epsilon}{2}}x^{-\frac{\alpha}{2}}\big)^{-b}\big]\dd y\dd x,\nonumber\\
	g_{b,1}(z)&=\int_{0}^{\infty}\int_{0}^{x} B_1^{-1}ze^{-zy}\frac{1-e^{-zxB_2/B_1}}{1-e^{-zx}}\nonumber\\
	&~~~~\times\big[1-\big(1+\theta S_\epsilon^{-\frac{\alpha \epsilon}{2}} z^{\frac{\alpha \epsilon}{2}} y^{\frac{\alpha \epsilon}{2}}x^{-\frac{\alpha}{2}}\big)^{-b}\big]\dd y\dd x, \nonumber
	\end{align} 
	$f_{b,2}(z)$ and $g_{b,2}(z)$ are given at the top of the next page.
	\label{thm:TFPCULMb}
\end{theorem}
\begin{IEEEproof}
	The expressions can be obtained by following similar steps as in the proof of Theorem~\ref{bthmoment}. A sanity check can be performed by letting $S_\epsilon\to\infty$ (\ie, $\hat p\to\infty$), which makes the second integral in \eqref{eq:TFPCULMb} zero and $f_{b,2}\to 0$, and thus recovers \eqref{eq:bthmoment}.	
\end{IEEEproof}
\begin{figure} [!t]
	\begin{center}
		\includegraphics[width=\figwidth]{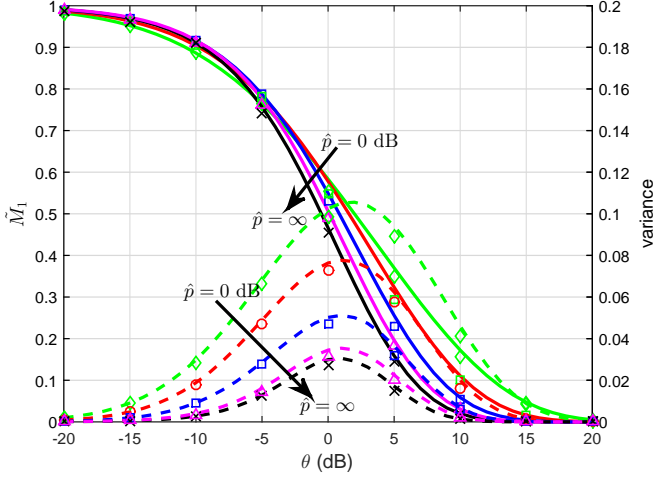}
		\caption{Uplink TFPC with $\lambda=0.1$, $\epsilon=1$, $\alpha=4$ and $\hat p = \{0,5,10,15,\infty\}$ (dB). The solid and dashed curves are the analytical results for $\tilde M_1$ and the variance, respectively. The markers are the simulation results.}
		\label{fig:UL_TFPC_M1Var_Lam0_1_Al4}
	\end{center}
\end{figure}
Figure \ref{fig:UL_TFPC_M1Var_Lam0_1_Al4} shows the analytical and simulation results of $M_1$ and the variance for TFPC with $\epsilon=1$, $\alpha=4$ and $\hat p = \{0,5,10,15\}$ (dB), the results of $\hat p = \infty$ are also shown. We can see that the curves of $\hat p =15$ dB are already very close to the non-truncated case $\hat p = \infty$.

\textbf{Remark 3}:
For the uplink with TFPC, we have the following scaling property.
\begin{enumerate}
	\item For a given $\theta$, $\lambda\to 0$ results in $\tilde{M}_b(\epsilon)\sim\tilde{M}_b(0)$, which can be verified through \eqref{eq:TFPCULMb} by noting that $\epsilon$ vanishes in $\lim\limits_{S_\epsilon\to 0}\tilde{M}_b$.
	This is consistent with our intuition since when the network density $\lambda$ is very low, the area of the Voronoi cell is large ($1/\lambda$ on average), resulting in a high probability for the link distance in each cell to be greater than $\hat{p}^{\frac{1}{\alpha\epsilon}}$. 
	\item From \eqref{eq:TFPCULMb} we can see that for TPFC, given $\alpha$, $\epsilon$ and $\theta$, $\lambda$ and $\hat p$ are connected through $S_\epsilon$, hence as long as $\lambda\hat p^{\delta/\epsilon}$ is kept constant, different combinations of the density and the maximum power yield the same results. In contrast, for FPC, $M_b$ and thus the meta distribution do not depend on $\lambda$. This gives insight about how to balance the network parameters for maintaing the performance for both the overall network and the individual users.  
\end{enumerate}  

\section{Meta Distribution for the Downlink}
\subsection{Network model}
In this section, we study the effect of fractional power control for downlink cellular networks using the framework of the meta distribution. We use the same model as for the uplink, namely the BSs are modeled by a homogeneous PPP $\Phib\subset\R^2$ with intensity $\lambda$, and the users follow the type I user point process. We assume the standard path loss law with the path loss exponent $\alpha > 2$ and power fading coefficients $\{h_x\}$, $x \in \Phib$ following exponential distribution with unit mean (\ie, Rayleigh fading) from BS $x$ to the typical user. Denote the link distance from the typical user to its serving BS $x_0 \in \Phib$ by $R$, the distance from the interfering BS $x$ to the typical user by $D_x$, and the interfering link distance of BS $x\in \Phi_{\rm b}\setminus \{x_0\}$ by $R_x$. Each BS $x$ uses fractional power control $P_x = R_x^{\alpha\epsilon}$, with the transmit power over unit distance normalized to $1$, and $\epsilon\in[0, 1]$ the compensation exponent. We focus on the typical user in the typical cell as in the uplink scenario. All the interfering BSs are further away than the serving BS of the typical user and are approximated by a PPP with the same intensity $\lambda$ seen by the typical user beyond the distance $R$. $R_x$ and $R$ are not independent but  identically distributed, and, in contrast to the uplink, $R_x$ is not bounded by $D_x$. Hence both of them have the same pdf $f_R(r) = \frac{5}{2}\pi\lambda r\exp\left(-\frac{5}{4}\lambda\pi r^2\right)$ in \eqref{eq:Rdistr2}, and, although correlated, they are assumed independent.

The SIR for the typical user is given by
\[\SIR \triangleq \frac{P_{x_0} h_{x_0}R^{-\alpha}}{\sum\limits_{x \in \Phi_{\rm b}\setminus \{x_0\}}P_x h_{x} D_x^{-\alpha}} = \frac{h_{x_0}R^{\alpha(\epsilon-1)}}{\sum\limits_{x \in \Phi_{\rm b}\setminus \{x_0\}}R_x^{\alpha \epsilon} h_{x} D_x^{-\alpha}}.\]

\subsection{Moments}
\begin{theorem}[Moments for the downlink with FPC]
	The $b$-th moment of the conditional success probability of the downlink Poisson cellular networks with FPC is closely approximated by
	\begin{equation}
	\tilde{M}_b = \int_{0}^{\infty} \exp\left(-z\left(\int_{1}^{\infty}f\left(z,x\right) \dd x +1\right)\right) \dd z,
	\label{eq:bthmomentDL}
	\end{equation}
	where $f(z,x)=\displaystyle\int_{0}^{\infty} B_1^{-1}ze^{-zy} \left(1- \left(1+\theta y^{\frac{\alpha \epsilon}{2}}x^{-\frac{\alpha}{2}}\right)^{-b} \right) \dd y$, and $B_1 = 5/4$.
	\label{bthmomentDL}
\end{theorem}

\begin{IEEEproof}
	The conditional coverage probability is
	\begin{align*}
	\Ps(\theta)&= \P\Big(h_{x_0} > \theta \sum_{x\in\Phi_{\rm b}\setminus\{x_0\}} h_xD_x^{-\alpha}R_x^{\alpha \epsilon}R^{\alpha(1-\epsilon)} \: \Big| \: \Phi_{\rm a},\Phi_{\rm b}\Big) \\
	&=\prod_{x\in\Phi_{\rm b}\setminus\{x_0\}} \frac{1}{1+\theta \frac{R^{\alpha(1-\epsilon)}R_x^{\alpha \epsilon}}{D_x^{\alpha}}}.
	\end{align*}
	\allowdisplaybreaks
	Then $M_b$ follows as
	\begin{align}
	M_b &= \mathbb{E}\prod\limits_{x \in \Phi_{\rm b}\setminus\{x_0\}} \frac{1}{\left(1+\theta \frac{R^{\alpha(1-\epsilon)}R_x^{\alpha \epsilon}}{D_x^{\alpha}}\right)^b} \nonumber \\
	&= \mathbb{E}\prod\limits_{x \in \Phi_{\rm b}\setminus\{x_0\}} \mathbb{E}_{R_x}\frac{1}{\left(1+\theta \frac{R^{\alpha(1-\epsilon)}R_x^{\alpha \epsilon}}{D_x^\alpha}\right)^b} \nonumber \\
	&\appa \mathbb{E}\prod\limits_{x \in \Phi_{\rm b}\setminus\{x_0\}} \int_{0}^{\infty} 2B_1\lambda\pi x e^{-B_1\lambda\pi x^2} \nonumber\\
	&~~~~\times\left(1+\theta x^{\alpha\epsilon} D_x^{-\alpha} R^{\alpha\left(1-\epsilon\right)}\right)^{-b} \dd x \nonumber \\
	&\eqb \mathbb{E}_R\exp\bigg(\int_R^\infty -2\lambda\pi a\Big(1-\int_0^\infty 2B_1\lambda\pi x e^{-B_1\lambda\pi x^2}\nonumber\\
	&~~~~\big(1+\theta x^{\alpha\epsilon} a^{-\alpha}R^{\alpha(1-\epsilon)}\big)^{-b} \dd x\Big) \dd a\bigg),
	\label{eq:Mb-FPC2}
	\end{align}	
	where (a) is due to the Rayleigh distribution approximation for $R_x$; (b) follows from the PGFL of the homogeneous PPP \cite{haenggi2012stochastic} and the fact that $D_x$ is strictly greater than $R$.	
	$\tilde{M}_b$ is then obtained by averaging over $R$ as
	\begin{align}
	&\tilde{M}_b \nonumber\\
	&= \int\limits_0^\infty 2B_1\lambda\pi r \exp\Bigg(\int\limits_r^\infty -2\lambda\pi a\bigg(1-\int\limits_0^\infty 2B_1\lambda\pi x e^{-B_1\lambda\pi x^2} \nonumber\\
	&~~~~\times\big(1+\theta x^{\alpha\epsilon} a^{-\alpha}r^{\alpha\left(1-\epsilon\right)}\big)^{-b} \dd x\bigg) \dd a \Bigg)e^{-B_1\lambda\pi r^2}\dd r.
	\label{eq:DLbthmomentexpanded}
	\end{align}
	Then using substitution $x/r = u$, $r/a = v$ and $e^{-B_1\lambda \pi r^2} = t$, after some simplification, we get the final result in \eqref{eq:bthmomentDL}. 
\end{IEEEproof}

\begin{figure} [t]
	\begin{center}
		\includegraphics[width=\figwidth]{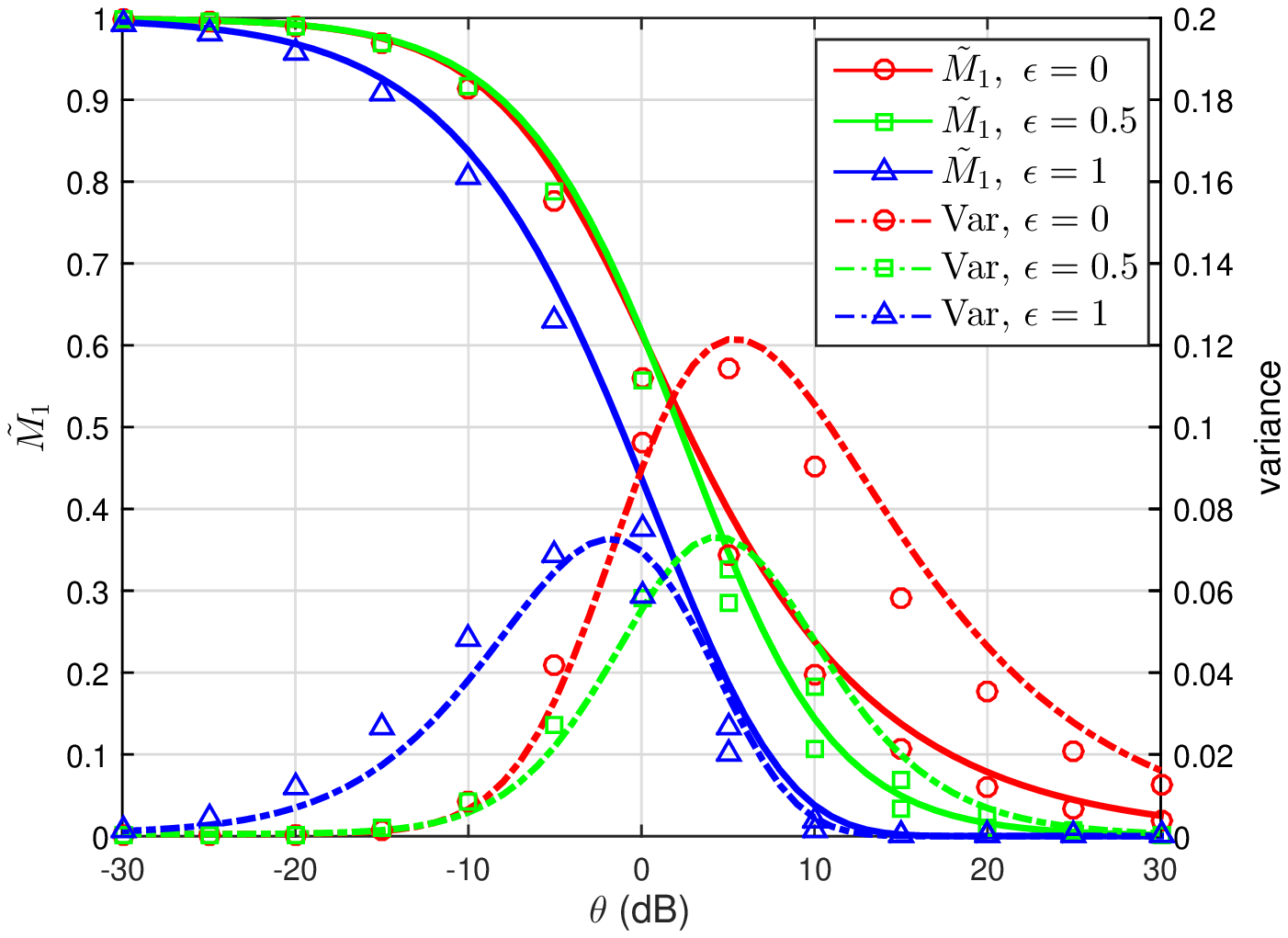}
		\caption{Downlink $M_1$ and variance of $P_s(\theta)$ obtained from simulation and analysis for different $\epsilon$ and $\alpha=4$. The curves are the analytical results from Theorem \ref{bthmomentDL} and the markers correspond to the simulation results.}
		\label{fig:DL-M1-Var1}
	\end{center}
\end{figure}
\begin{figure} [t]
	\begin{center}
		\includegraphics[width=\figwidth]{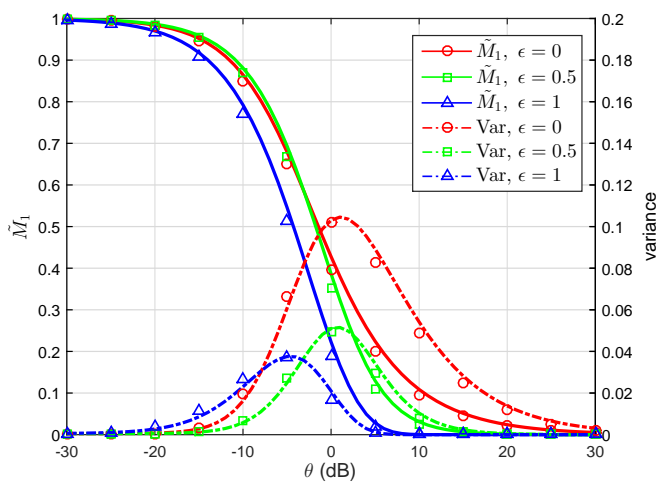}
		\caption{Downlink $M_1$ and variance of $P_s(\theta)$ obtained from simulation and analysis for different $\epsilon$ and $\alpha=3$. The curves are the analytical results from Theorem \ref{bthmomentDL} and the markers correspond to the simulation results.}
		\label{fig:DL-M1-Var2}
	\end{center}
\end{figure}

\figref{fig:DL-M1-Var1} and \figref{fig:DL-M1-Var2} show the downlink standard success probability $M_1 = p_{\rm s}$ and the variance of the conditional success probability as a function of $\theta$ for FPC exponents $\epsilon = 0, 0.5, 1$ and $\alpha = 4$ and $\alpha = 3$, respectively. The solid and dashed curves correspond to the simulation result, and the markers correspond to the analytical result by Theorem \ref{bthmomentDL}. These results show that the analytical results closely match the simulation results and also reveal that an appropriate FPC exponent can help mitigate the variance while maintaining the level of $M_1$, at least for a certain range of $\theta$. For example, for $\theta$ ranging from $-10$ dB to $0$ dB, $\epsilon = 0.5$ performs nearly the same as $\epsilon = 0$ in terms of $M_1$, however the former reduces the variance significantly compared to the latter. Hence, by studying power control with the tool of meta distribution, we find that FPC can bring some fairness benefits to the downlink. 

Based on Theorem \ref{bthmomentDL}, we have the following statements for the downlink mean local delay.
\begin{corollary}[Mean local delay for the downlink]
	\label{cor:MLD_DL}
	When $\alpha>2$, the downlink mean local delay for Poisson cellular networks with FPC is closely approximated by 
	\begin{equation} \label{eq:MLD_DL} 
	\tilde{M}_{-1}	= \int_0^\infty\exp(c\Gamma(1+\rho) y^{1-\rho} -y)\dd y,
	\end{equation}
	where $c=B_1^{-1}\theta\delta/(1-\delta)$, $\rho=\epsilon/\delta$ and $\delta=2/\alpha$.
\end{corollary}
\begin{IEEEproof}
	By substituting $b=-1$ in \eqref{eq:bthmomentDL}, we obtain $\tilde{M}_{-1}$ in the form
	\begin{align*}
	&\tilde{M}_{-1}\nonumber\\
	&= \int_0^\infty \exp\bigg(\frac{\theta z^2}{B_1}\int_1^\infty\int_0^\infty e^{-zy}y^{\alpha\epsilon/2} x^{-\alpha/2}\dd y \dd x -z \bigg) \dd z \nonumber \\
	&= \int_0^\infty \exp\bigg(\frac{\theta}{B_1} \int_1^\infty x^{-\alpha/2}\dd x \int_0^\infty z^2 e^{-zy}y^{\alpha\epsilon/2}\dd y -z \bigg) \dd z,
	\end{align*}
	by noticing that $\int_1^\infty x^{-\alpha/2}\dd x = \frac{1}{-1+\alpha/2}$ and $\int_0^\infty z^2e^{-zy}y^{\alpha\epsilon/2}\dd y = z^{1-\epsilon\alpha/2}\Gamma(1+\epsilon\alpha/2)$, after the substitution, we get the final expression in \eqref{eq:MLD_DL}.
\end{IEEEproof}	 
  
\begin{corollary}[Convergence of downlink mean local delay]
	\label{cor:MLD_DL2}
	When $\alpha > 2$, $\tilde{M}_{-1}$ is finite for any $\theta$ if $0<\epsilon\leq\delta$; for $\epsilon=0$, $\tilde{M}_{-1}$ is finite if $\theta<B_1(1/\delta-1)$; if $\epsilon>\delta$, $\tilde{M}_{-1}$ is $\infty$ for all $\theta$. 
\end{corollary}
\begin{IEEEproof}
	In \eqref{eq:MLD_DL}, since $\delta<1$, we have $c>0$ and $1-\rho\leq1$. To make $\tilde{M}_{-1}$ finite, $1-\rho$ must be non-negative. Thus we have $0\leq 1-\epsilon/\delta\leq1$, which gives $0\leq \epsilon \leq\delta$. Morever, when $\epsilon=0$, then $\rho=0$, $c$ must be smaller than 1 to guarantee the convergence of the integral.  
\end{IEEEproof}	

{\textbf{Remark 4:} Cor.~\ref{cor:MLD_DL2} indicates that the phase transition of the mean local delay in downlink Poisson cellular networks only occurs at $\epsilon=0$, \ie, when there is no power control. The expression for $\tilde{M}_{-1}$ at $\epsilon=0$ is given by 
\begin{equation}
\tilde{M}_{-1} = \frac{1-\delta}{1-\delta(1+B_1^{-1}\theta)} = \frac{1}{1-\frac{\delta}{B_1(1-\delta)}\theta}, ~\theta< B_1/\delta-B_1.
\label{eq:Mbeq0}
\end{equation}  
The critical SIR threshold is $\theta_{\rm c} = B_1(1/\delta-1)$.}

{\textbf{Remark 5:} For $\epsilon=\delta/2$ and $\epsilon=\delta$, $\tilde{M}_{-1}$ also admits closed-form expressions, given by \eqref{eq:Mbeq1} and \eqref{eq:Mbeq2}, respectively.
\begin{align}
\label{eq:Mbeq1} \epsilon = \frac{\delta}{2}:\quad &\tilde{M}_{-1} = 1+\frac{\theta}{2 B_1}\cdot\frac{\delta\sqrt{\pi}}{1-\delta}\exp\left(\frac{\theta^2}{4 B_1^2}\cdot\left(\frac{\delta}{1-\delta}\right)^2\right)\nonumber\\
&~~~~\times\erfc\left(-\frac{\theta}{2 B_1}\cdot\frac{\delta}{1-\delta}\right) \\
\label{eq:Mbeq2} \epsilon = \delta:\quad &\tilde{M}_{-1} = \exp\left(\frac{\delta}{B_1(1-\delta)}\theta\right)
\end{align}
	
Interestingly, $\frac{\delta}{1-\delta}$ is exactly the mean interference-to-signal ratio (MISR) of the PPP introduced in \cite{Haenggi14wcl}, hence, \eqref{eq:Mbeq0}, \eqref{eq:Mbeq1} and \eqref{eq:Mbeq2} can also be expressed as a function of the MISR.}
\begin{figure} [t]
	\begin{center}
		\includegraphics[width=\figwidth]{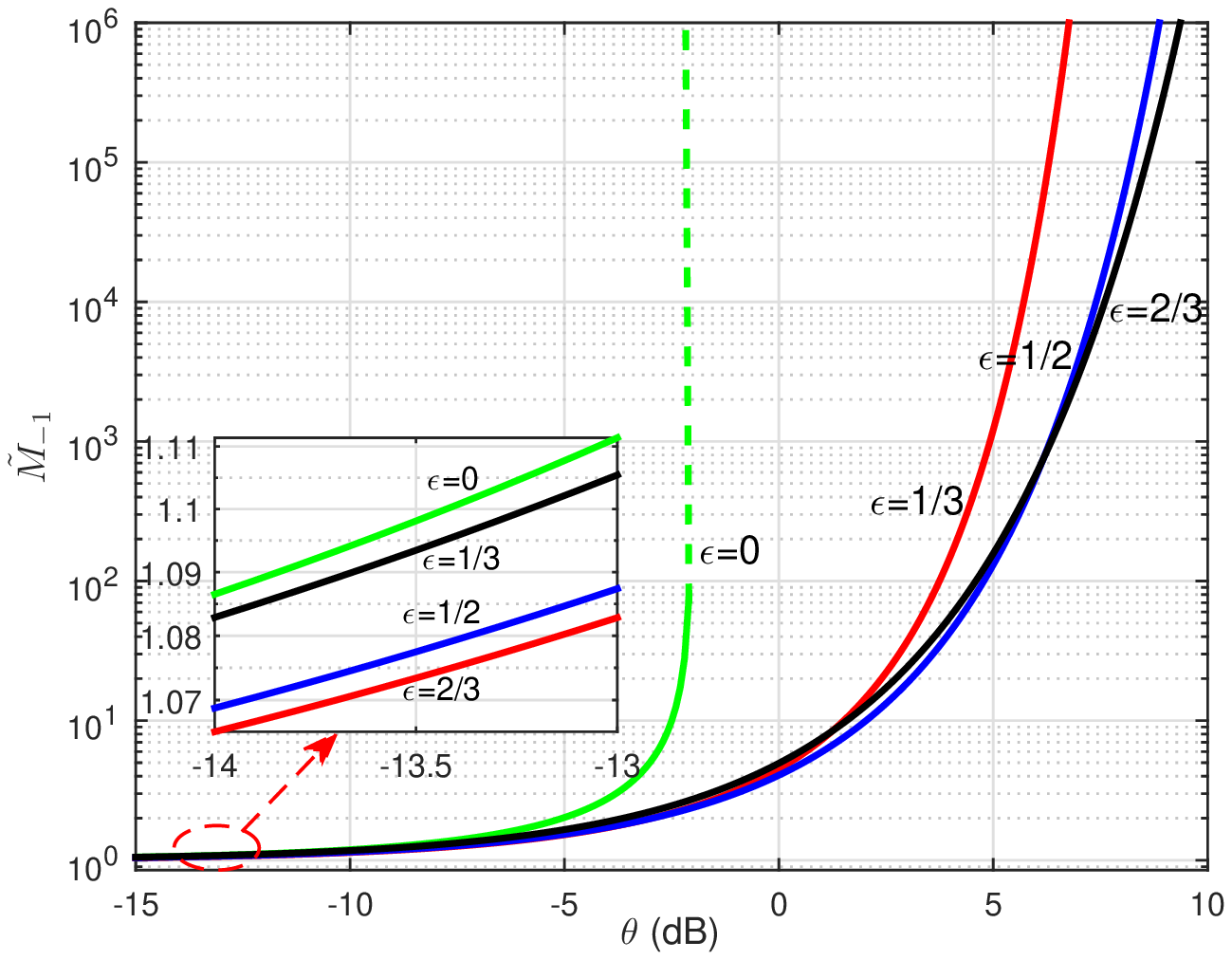}
		\caption{Downlink mean local delay as a function of $\theta$ obtained from analytical results for $\alpha=3$ and different $\epsilon$, where $\epsilon=0$ corresponds to \eqref{eq:Mbeq0}, $\epsilon=1/3$ corresponds to \eqref{eq:Mbeq1} and $\epsilon=2/3$ corresponds to \eqref{eq:Mbeq2}. A phase transition occurs at $\theta\approx-2$ dB when $\epsilon=0$.}
		\label{fig:DL_MeanLocalDelay_al3}
	\end{center}
\end{figure}
\begin{figure} [t]
	\begin{center}
		\includegraphics[width=\figwidth]{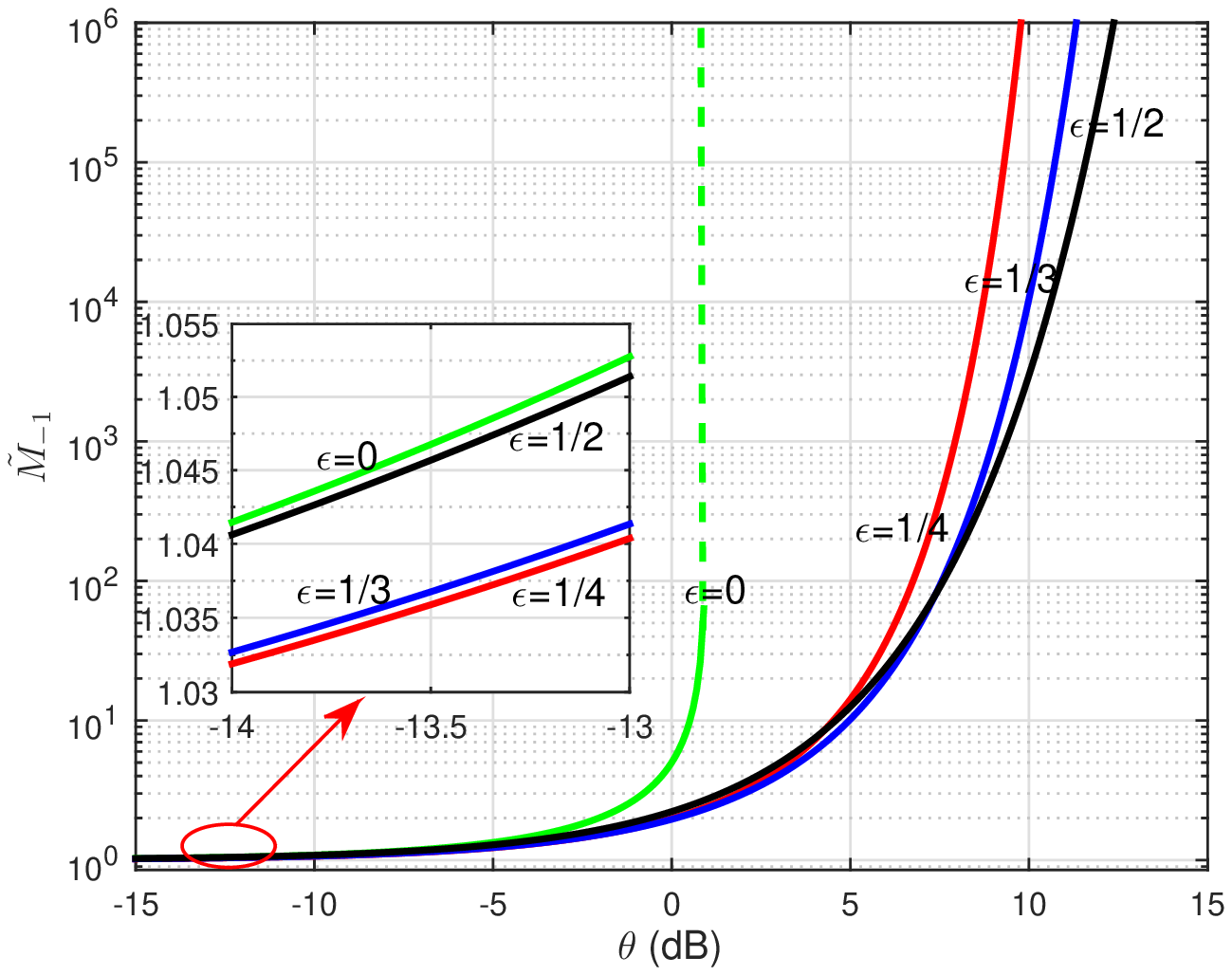}
		\caption{Downlink mean local delay as a function of $\theta$ obtained from analytical results for $\alpha=4$ and different $\epsilon$, where $\epsilon=0$ corresponds to \eqref{eq:Mbeq0}, $\epsilon=1/4$ corresponds to \eqref{eq:Mbeq1} and $\epsilon=1/2$ corresponds to \eqref{eq:Mbeq2}. A phase transition occurs at $\theta\approx1$ dB when $\epsilon=0$.}
		\label{fig:DL_MeanLocalDelay_al4}
	\end{center}
\end{figure}

\figref{fig:DL_MeanLocalDelay_al3} and \figref{fig:DL_MeanLocalDelay_al4} show the analytical results of $\tilde{M}_{-1}$ for different $\epsilon$. It can be seen that for large $\theta$, $\epsilon=\delta$ is optimal for the minimization of the mean local delay, while for small $\theta$, $\epsilon=\delta/2$ is optimal to minimize the mean local delay.

{\textbf{Remark 6:}~Letting $\rho_{\rm opt}(c) = \arg\min\limits_\rho \tilde{M}_{-1}$, a detailed numerical study of \eqref{eq:MLD_DL} shows that $\lim\limits_{c\to 0} \rho_{\rm opt} = 1/2$ and $\lim\limits_{c\to \infty} \rho_{\rm opt} = 1$. A small $c$ can be achieved by either a small $\theta$ or a small $\delta$, while a large $c$ can be achieved by either a large $\theta$ or a $\delta$ close to $1$.}

\subsection{Meta distribution: analytical curves, and beta approximation}
\begin{figure} [t]
	\begin{center}
		\includegraphics[width=\figwidth]{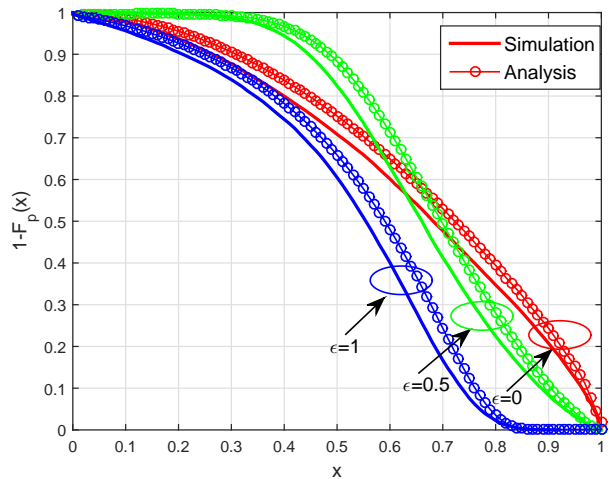}
		\caption{Both the simulated and the analytical meta distribution \eqref{exact} for $\alpha=3$, $\theta = -5$ dB in downlink Poisson cellular networks.  }
		\label{fig:DLMetaThmSim}
	\end{center}
\end{figure}

\begin{figure} [t]
	\begin{center}
		\includegraphics[width=\figwidth]{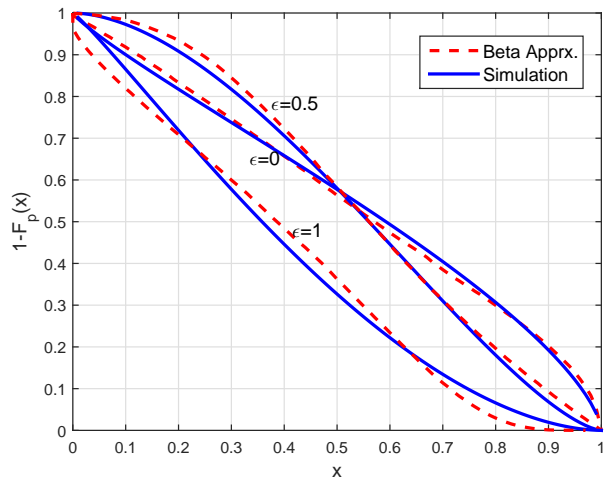}
		\caption{The meta distribution (simulation result) and the beta distribution approximation for $\alpha=4$, $\theta = 0$ dB and $\epsilon \in \{0, 0.5, 1\}$ in the downlink.}
		\label{fig:DLBetaApprxMeta}
	\end{center}
\end{figure}

Following the same methods as in the uplink analysis, the analytical results of the downlink meta distribution for Poisson cellular networks can be calculated by using the Gil-Pelaez theorem, which are shown in \figref{fig:DLMetaThmSim}. The beta approximation of the meta distribution through matching the mean and variance is shown in \figref{fig:DLBetaApprxMeta}. 


\section{Conclusions}
This paper applies the concept of the meta distribution, which is the distribution of the conditional success probability given the point process, to study the uplink and downlink scenarios in Poisson cellular networks with fractional power control. For the uplink scenario, the interfering user point process relative to the typical BS is approximated by a non-homogeneous PPP whose intensity function was obtained by analyzing the pair correlation function of the interfering points in \cite{2016arXivUserHaenggi}. This approximation yields analytical results that are highly consistent with the simulation results. For both scenarios, the general expression of the moments, the analytical results for the mean local delay and the analytical expression for the distribution of the conditional success probability are provided, revealing much more detailed information about the user performace in cellular networks than just the standard success probability $M_1$. Moreover, it is shown that both the uplink and downlink meta distributions can be closely approximated by the beta distribution through matching the mean and variance, which is convenient for the network performance analysis with no need to calculate higher-order moments. Further, the truncated fractional power control model as an extension for the FPC model with a maximum transmit power constraint is also evaluated for the uplink scenario.  

The study of the mean local delay does not indicate any phase transition in the uplink. For the downlink, a phase transition only occurs when there is no power control ($\epsilon=0$). The optimum power control parameter that minimizes the mean local delay in the downlink is between $\delta/2$ and $\delta$.

The investigation of the fractional power control in both the uplink and downlink shows that compensating the path loss sensibly improves the user fairness while maintaining the average overall network performance. It also reveals that the effect of FPC is mainly a {\em{concentration}} in the user performance levels, which means $M_1$ alone does not give enough information about what parameter to use, but by virtue of the meta distribution, we can obtain more fine-grained information (e.g., the variance and the 5\% user performance) to find the best $\epsilon$. Such advantages of the meta distribution are helpful to the operators in practical network deployment and configuration.

\bibliographystyle{IEEEtran}

\end{document}